\documentclass[12pt]{article}
\usepackage{amsmath}
\usepackage{amssymb}
\usepackage{epsfig}
\usepackage{color}

\def\bm{\boldsymbol}
\newcommand{\bea}{\begin{eqnarray}}
\newcommand{\eea}{\end{eqnarray}}
\newcommand{\be}{\begin{eqnarray}}
\newcommand{\ee}{\end{eqnarray}}
\newcommand{\no}{\nonumber \\}
\newcommand{\etal}{{\it et al.}~}

\def\H#1{{}^{#1}\mbox{H}}
\def\He#1{{}^{#1}\mbox{He}}
\def\MeV{{\mbox{MeV}}}
\def\gA{g_A}
\def\fpi{f_\pi}

\def\ttz{(\tau_1\times\tau_2)^z}

\def\fpi{f_\pi}

\def\TminusS{{{\hat T}_S^{(-)}}}
\def\TminusT{{{\hat T}_T^{(-)}}}
\def\TplusS{{{\hat T}_S^{(+)}}}
\def\TplusT{{{\hat T}_T^{(+)}}}

\def\TtimesS{{{\hat T}_S^{(\times)}}}
\def\TtimesT{{{\hat T}_T^{(\times)}}}

\def\vR{{\bm R}}

\def\vr{{\bm r}}
\def\vq{{\bm q}}
\def\vk{{\bm k}}
\def\vp{{\bm p}}
\def\vbp{{\bar {\bm p}}}

\def\vs{{\bm \sigma}}
\def\vmu{{\bm \mu}}

\def\hatr{{\hat \vr}}
\def\nlo#1{\mbox{N$^{#1}$LO}}

\begin{document}
\vskip 0.4cm \hfill {\bf } \vskip 1cm
\begin{center}
{\LARGE Effective field theory approach for the M1 properties of
A=2 and 3 nuclei}
\vskip 1cm {\large Young-Ho
Song$^{(a)}$\footnote{E-mail:singer@phya.snu.ac.kr}, Rimantas
Lazauskas $^{(b)}$\footnote{E-mail:lazauskas@lpsc.in2p3.fr},
Tae-Sun Park$^{(c)}$\footnote{E-mail: tspark@kias.re.kr}\footnote{
Present address: Department of Physics and
Basic Atomic Energy Research Institute, Sungkyunkwan University,
Suwon 440-746, Korea },
Dong-Pil Min$^{(a)}$\footnote{E-mail: dpmin@snu.ac.kr} }
\end{center}
\vskip 0.5cm
\begin{center}
$^{(a)}$ {\it School of Physics, Seoul National University, Seoul
151-742, Korea}

$^{(b)}$ {\it Institut de Physique Nucl\'eaire,
F-91406 Orsay cedex, France}

$^{(c)}$ {\it School of Physics,
Korea Institute for Advanced Study, Seoul 130-012, Korea}
\end{center}
\vskip 0.5cm
\begin{abstract}

The magnetic moments of ${}^2\mbox{H}$,
$\He3$ and $\H3$ as well as the thermal
neutron capture rate on the proton are calculated
using heavy baryon chiral perturbation theory {\it \`{a} la} Weinberg.
The M1 operators have been derived up to \mbox{N$^3$LO}. The
nuclear matrix elements are evaluated with the use of wave
functions obtained by carrying out variational Monte Carlo
calculations for a realistic nuclear Hamiltonian involving
high-precision phenomenological potentials like Argonne
Av18 and Urbana IX tri-nucleon interactions. We discuss the
potential- and cutoff-dependence of the results.

%--------------------------------------------------------------------
\end{abstract}

\newpage
%%%%%%%%%%%%%%%%%%%%%%%%%%%%%%%%%%%%%%%%%%%%%%%%%%%%%%%%%%%%%%%%%%%%%%%%%

\section{Introduction}
\indent

The standard nuclear physics approach (SNPA) based on
meson-exchange currents\cite{MEC},
high-precision phenomenological potentials and
the state-of-the-art techniques
for obtaining a few-body nuclear wave functions
has achieved tremendous
progress in understanding light nuclear systems\cite{SNPA}.
SNPA is particularly useful
when the impulse contributions
and the well-known long-ranged
one-pion-exchanges predominate,
as was formulated in terms of the so-called
{\em chiral filtering}~\cite{KDR}.
However, if more complicated meson exchange processes are
important, SNPA becomes model-dependent
because the phenomenological description
of the nuclear potentials and transition operators
does not allow the unique description
of their short-range behavior.

The merit of effective field theory (EFT)~\cite{review:eft} is that it
offers a systematic way to control short-range physics,
which in principle allows more accurate
theoretical predictions
than achievable in the conventional approach
~\cite{predict}.
An example is the $hep$ process,
$\He3+p\rightarrow \He4+\nu_e+e^+$,
which is potentially important
for solar neutrino physics.
A reliable estimation of
the $S$-factor for this reaction
has been a long-standing challenge in nuclear physics~\cite{challenge};
its theoretical estimates have varied by orders of
magnitude in the literature.
A highly elaborate SNPA calculation of this S-factor
was carried out
by Marcucci {\it et. al.}~\cite{Marcucci:2000bh},
%$S(0)=9.64\times 10^{-20}\ \mbox{KeV-b}$,
but it was a recent EFT-based calculation
by Park {\it et al.}~\cite{hep} that
provided definitive support to the SNPA results,
giving in addition a quantitative error estimate.
In \cite{hep}, the authors have developed an EFT approach,
called EFT$^*$ or
MEEFT ({\em more effective} effective field theory)
and performed a parameter-free
calculation of the $hep$ cross section,
which has led to an estimate of the $S$-factor
with $\sim$15\% precision;
for a detailed review, see Ref.~\cite{KP}.
The same method has marked successes also in describing the highly
suppressed isoscalar amplitude pertinent to the $n\!+\!p\!\rightarrow\!
d\!+\!\gamma$ process~\cite{excV,isoS}, as well as the weak processes
like muon capture on the deuteron~\cite{mu-d} and $\nu$-$d$
scattering~\cite{nu-d1,nu-d2}. For a recent review,
see Ref.~\cite{kubodera}.

The strategy of MEEFT,
as explained in \cite{KP,hep,park2001}, is in the
spirit of the original Weinberg scheme~\cite{weinberg}
based on the chiral
expansion of {\it irreducible} terms.
In MEEFT the relevant current operators
are derived systematically by applying HB$\chi$PT
to a specified order.
Nuclear matrix elements corresponding to
these current operators are evaluated
with the use of realistic nuclear wave functions
obtained by applying an ab-initio or quasi-ab-initio few-body
calculation method to a realistic phenomenological nuclear Hamiltonian
that involves high-precision phenomenological potentials.
In MEEFT, as in the usual EFT, short-range physics is
simulated by contact counter-terms and the coefficients of these
terms, called the low-energy constants (LECs), are determined by
requiring that a selected set of experimental data is reproduced.
This renormalization procedure is expected to
remove, to a large extent,
model-dependence that might creep in
through the phenomenological
parametrization of
short-range physics in the adopted nuclear interaction.
We note that the strategy employed here is closely related to the
one used to construct a universal $V_{low-k}$
in the recently developed
renormalization-group approach to nuclear
interactions~\cite{stonybrook}.

The purpose of this letter is to demonstrate that MEEFT provides a
practical and reliable tool to compute
the low energy
%electromagnetic
M1 properties of few-nucleon systems in a
largely model-independent way.
We calculate here four low-energy
electromagnetic observables, namely
the magnetic moments of $\H2$, $\He3$ and $\H3$ and the
rate of radiative capture of a thermal neutron on a proton.
We derive all the relevant operators
up to the next-to-next-to-next leading order (\nlo3) of the
perturbation series, which include the short-range contact terms.
We use the experimental values of two
of the above-mentioned four observables to fix the coefficients
of the contact terms appearing
%in MEEFT
at \nlo3 and make
%the {\it postdiction}
prediction for the remaining two observables.
We test the
consistency as well as the model-independence of the obtained results.
To our knowledge, this is the first EFT calculation of
the magnetic moments of the A=3 systems.
%$\mu(\H3)$ and $\mu(\He3)$ yet.

Before closing this section, we remark that
other EFT methods can also be adopted, at least in principle,
for studying the M1 properties of few-nucleon systems
at low energy.
One of such possible alternatives consists in
deriving not only the transition operators
but also the nuclear interactions
using the same EFT.
This method
has the advantage of being more transparent
in discussing the order counting and model-independence;
for the recent developments
in the EFT studies of nuclear
potentials, see Refs.~\cite{egm1,egm2,epel}
and references therein.
Such a study is expected to be highly illuminating
particularly for the electric transitions
because gauge invariance (or charge conservation)
is automatically guaranteed order by order.
Another possibility is to use the so-called
pionless EFT, where even the pions are integrated out
leaving only the nucleon field as a pertinent degree of freedom.
Pionless EFT has been successfully applied to
some processes involving two- or three-nucleon systems~\cite{pionless}.

%%%%%%%%%%%%%%%%%%%%%%%%%%%%%%%%%%%%%%%%%%%%%%%%%%%%%%%%%%%%%%%%%%%%%%%%%%%%%%

\section{Current operators}
\indent

The relevant operator for the magnetic moment and
radiative $np$ capture at threshold is the M1 operator $\vmu(q)$,
\be
\label{M1}
\vmu(q)\equiv \left(\frac{iq}{\sqrt{6\pi}}\right)^{-1}{\hat T}^{Mag}_{10}(q)
\ee
where ${\hat T}^{Mag}_{10}(q)$ is defined in \cite{Walecka},
$q^\mu=(\omega,\ \vq)$ is the momentum carried out by the photon
($q^\mu=0$ for the magnetic moment
while $q^\mu \neq 0$ for $np$ capture), and $q\equiv|\vq|$.

We derive the M1 operator in HB$\chi$PT,
which contains the nucleons and pions as pertinent degrees of
freedom with all other massive fields integrated out. In
HB$\chi$PT the electromagnetic currents and M1 operator are expanded
systematically with increasing powers of $Q/\Lambda_\chi$, where
$Q$ stands for the typical momentum scale of the process and/or
the pion mass, and $\Lambda_\chi\sim 4\pi f_\pi \sim m \sim 1$~GeV
is the chiral scale, $f_\pi\simeq 92.4$ MeV is the pion decay
constant, and $m$ is the nucleon mass.
%\footnote{\syh{We can count $m_N\sim \frac{\Lambda_\chi^2}{Q}$
%according to
%Weinberg\cite{weinberg}. The result is not sensitive to the treatment of
%$m_N$}}.
We remark
that, while the nucleon momentum $\vp_i$ is of order of $Q$,
its energy ($\sim \vp_i^2/m$) is  of order of
$Q^2/m$, and consequently the four-momentum of the emitted photon
$q^\mu$ (with $\omega=|\vq|$)
should also be counted as ${\cal O}(Q^2/m)$.
%\footnote{
%This counting is specific to real photons where $\omega=|\vq|$.
%For virtual photons, $|\vq|$ should be in general counted as
%${\cal O}(Q)$ while $\omega$ remains to be counted as ${\cal O}(Q^2/m)$.}
%
In this work we include {\em all} the contributions up to \nlo3,
where $\nlo{\nu}$ denotes terms of order of $(Q/\Lambda_\chi)^\nu$
compared to the leading one-body contribution.
It is worth mentioning that
there is a different power counting scheme where
the nucleon mass is regarded as heavier than the chiral scale,
see Refs.~\cite{egm1,egm2} for details.
However, the use of this alternative counting scheme
would not affect the results to be reported in this article
since the difference
between the two counting schemes
would appear only at orders higher
than explicitly considered here (\nlo3).

The one-body (1B) M1 operator including the relativistic
corrections reads
\be
\vmu_{\rm 1B}(q) &=& \sum_i \frac{1}{2 m_p} \Big\{
  \hat j_0(q r_i) \left[
 \vs_i \left( \mu_i - Q_i \frac{\vbp_i^2}{2 m^2}\right)
  -\frac{\mu_i-Q_i}{2 m^2}  \bar{\vp}_i\vs_i\cdot \bar{\vp}_i \right]
 \nonumber \\
 &+& \hat j_1(q r_i) \left[
 Q_i \vr_i\times \vbp_i \left(1 - \frac{\vbp_i^2}{2 m^2}\right)
 -\frac{w(2\mu_i-Q_i)}{4m} i\vr_i \times (\bar{\vp}_i\times\vs_i) \right]
 \nonumber\\
 &+& \frac{(q r_i)^2}{30} \hat j_2(q r_i)
  (3 \hat \vr_i \,\hat \vr_i\cdot \vs_i - \vs_i)
 + \cdots \Big\}
\ee
where $\hat j_n(x)\equiv \frac{(2n+1)!!}{x^n} j_n(x)= 1 +
{\cal O}(x^2)$, $Q_i=(1+\tau^z_i)/{2}$
is the charge of the $i$-th nucleon,
and $\mu_i=(\mu_s+\tau^z_i
\mu_v)/{2}$ is the magnetic moment in units of the nuclear magneton,
$\mu_N=e/(2 m_p)$,
with $\mu_s=\mu_p+\mu_n\simeq 0.8798$ and
$\mu_v=\mu_p-\mu_n\simeq 4.7059$; $\vbp_i\equiv
\frac{1}{2}(i\stackrel{\leftarrow}{\nabla}_i
-i\stackrel{\rightarrow}{\nabla}_i)$
with the understanding
that the derivatives act only on the wave functions,
and $r_i\equiv |\vr_i|$.
In the above equation
the familiar $(\mu_i \vs_i+ Q_i\vr_i\times
\bar{\vp}_i)$ term is of leading order (LO),
while all the others terms are $\nlo{2}$.
Since there are  no $\nlo{3}$ contributions to the
$\vmu_{\rm 1B}$, the neglected terms are of \nlo4 and higher
orders.

Corrections to the 1B operator are due to the meson-exchange
currents (MEC). Up to \nlo3, only two-body (2B) contributions enter;
three-body (3B) currents are \nlo4 or higher order.
It is to be emphasized that MECs derived in EFT
are meaningful only up to a certain momentum scale $\Lambda$,
where $\Lambda$ is the cutoff below which the chosen explicit
degrees of freedom reside. This cutoff may be realized by
introducing a Gaussian regulator
%\syh{,$S_\Lambda(\vk)=e^{-\frac{\vk^2}{2\Lambda}}$,}
in performing the Fourier
transformation of the MECs from momentum space to coordinate space
\cite{hep}.
It is to be noted that the contributions due to high momentum
exchanges (above the cutoff scale)
are not simply ignored but,
as we will discuss later,
they are accounted for by the renormalization of the
contact-term coefficients.

We decompose the two-body current into the
{\em soft}-one-pion-exchange ($1\pi$) term, vertex corrections to the
one-pion exchange $(1\pi C)$ term, the so-called {\it fixed} term
contribution $(1\pi:fixed)$, the two-pion-exchanges $(2\pi)$ term, and
the contact-term contribution $(CT)$,
\be \vmu_{\rm 2B}(q) = \sum_{i<j} \left[\vmu_{ij}^{1\pi} +
\left(\vmu_{ij}^{1\pi C} + \vmu_{ij}^{1\pi:fixed} +
\vmu_{ij}^{2\pi} + \vmu_{ij}^{CT}\right)\right] = \nlo{} + \nlo3.
\ee

\begin{figure}
\begin{center}
\label{fig:1pi} \epsfig{file=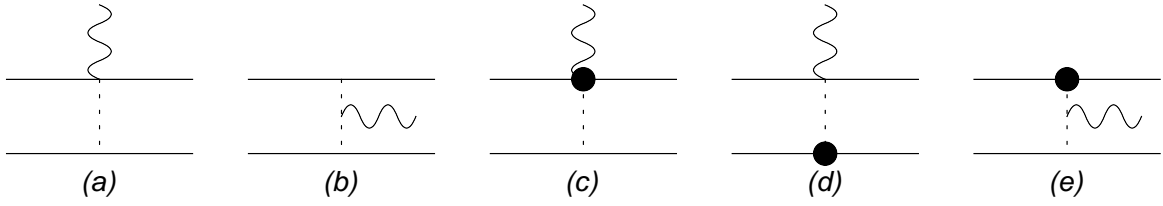}
\caption{Tree graphs of NLO and \nlo3.
One-pion exchange ``seagull"(a) and ``pion-pole" diagram
(b) contribute to the $\mu^{1\pi}$.
Diagrams (c)-(e) contribute to the
$\mu^{1\pi C}$ and $\mu^{1\pi;fixed}$ at \nlo3.
The dot represents
the vertex corrections coming from NLO or \nlo2 lagrangian.}
\end{center}
\end{figure}

\begin{figure}
\begin{center}
\label{fig:N3LO-CT} \epsfig{file=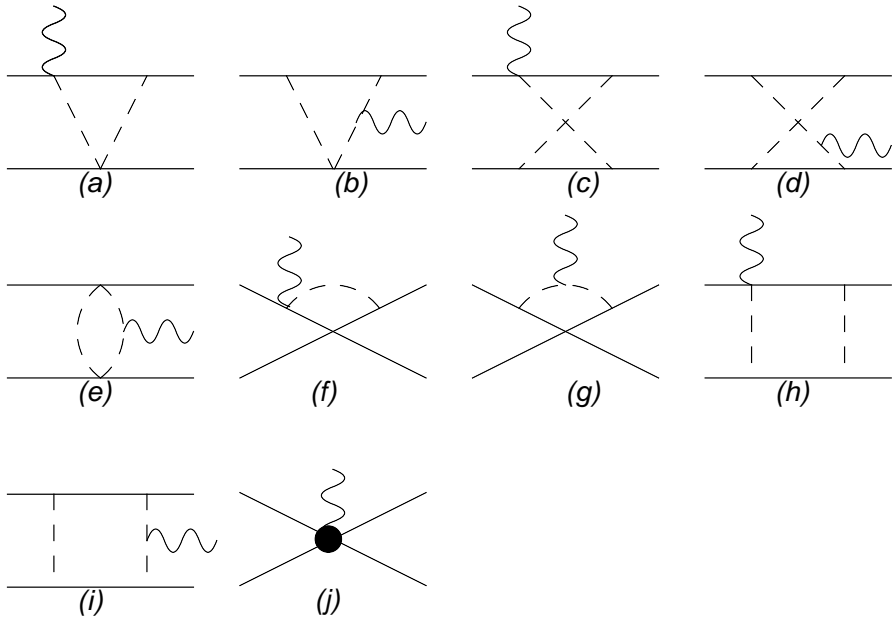}
\caption{Diagrams
contributing to $\mu^{2\pi}$ (a)-(i) and $\mu^{CT}$ (j) at \nlo3.}
\end{center}
\end{figure}
\noindent
The leading MEC due to the {\em soft}-one-pion-exchange
($1\pi$) is NLO and given as
 \be \vmu_{12}^{1\pi} &=& \frac{g_A^2}{8f_\pi^2}
  \left[{\hat T}_{S}^{(\times)}
  \left(\frac{2}{3}y_{1\Lambda}^\pi(r)-y_{0\Lambda}^\pi(r)\right)
        -{\hat T}_{T}^{(\times)}y_{1\Lambda}^\pi(r)\right]
   \hat j_0 (q R)
\no &-&\frac{g_A^2 m_\pi^2}{24 f_\pi^2}\tau_{\times}^z
     \vR \times \vr
          \left[\vs_1\cdot\vs_2 {\bar y}_{0\Lambda}^\pi(r)
           +S_{12}y_{2\Lambda}^\pi(r)\right]
          \hat j_1 (q R)
      + \cdots,
\label{1pi}
\ee
where $g_A\simeq 1.2695$, $\vr=\vr_1-\vr_2$, $r=|\vr|$,
$\vR=(\vr_1+\vr_2)/2$,
$R=|\vR|$, and
\bea
\delta_\Lambda(r)&=&\int\!\! \frac{d^3 \vk}{(2\pi)^3}
\,e^{-\vk^2/\Lambda^2}\,
e^{i\vk\cdot\vr},
\no
y_{0\Lambda}^\pi(r)&=&\int \frac{d^3 \vk}{(2\pi)^3}
\,e^{-\vk^2/\Lambda^2}\,
e^{i\vk\cdot\vr}
\frac{1}{\vk^2+m_\pi^2},
\eea
$y_{1\Lambda}^\pi=-r\frac{d}{dr}y_{0\Lambda}^\pi$,
$y_{2\Lambda}^\pi=\frac{r}{m_\pi^2}
\frac{d}{dr}\frac{1}{r}\frac{d}{dr} y_{0\Lambda}$
and
$S_{12}=3\vs_1\cdot\hatr\vs_2\cdot\hatr-\vs_1\cdot\vs_2$.
We have
also defined ${\hat T}_{S}^{(\odot)}\equiv \tau_\odot^z
\vs_\odot$ and ${\hat T}_{T}^{(\odot)}\equiv \tau_\odot^z
  \left[\hatr\,\hatr\cdot\vs_\odot-\frac13\vs_\odot\right]$,
$\tau_{\odot}=\tau_1\odot\tau_2$, $\vs_{\odot}=\vs_1\odot\vs_2$,
where $\odot= \pm,\, \times$.
Strictly speaking, $\vmu_{ij}^{1\pi}$ contains not only NLO
but also admixtures of \nlo3 and higher order terms;
if we expand
$\vmu_{ij}^{1\pi}$ in powers of $q$, the
$q$-independent part is NLO while the remaining $q$-dependent part
is \nlo3 (or of even higher orders). Furthermore, if $q$ is
non-zero, as is the case for the $np$ capture, Fourier
transformation for the {\em soft}-one-pion-exchange current
becomes rather involved, generating complicated \nlo3
contributions, which are denoted by the ellipsis in the above
equation and will be reported elsewhere~\cite{elsewhere}.
A numerical evaluation
shows that the contribution of the \nlo3 part
in the $1\pi$ current is
negligibly small.

There are no corrections at \nlo2,
and all the contributions up to \nlo3 are included in
this work.
Among the \nlo3 contributions,
the one-loop vertex correction of the one-pion
exchange
has been investigated in detail in Ref.~\cite{excV,isoS},
\be
\vmu_{12}^{1\pi C} &=& -\frac{\gA^2}{8 \fpi^2}
  ({\bar c}_\omega+{\bar c}_\Delta)
  \left[({\hat T}_{S}^{(+)}+ {\hat T}_{S}^{(-)})
  \frac{{\bar y}^\pi_{0\Lambda}}{3}
  +({\hat T}_{T}^{(+)} + {\hat T}_{T}^{(-)})\,y^\pi_{2\Lambda}\right]
  \hat{j}_1(qR)
\no
  &&
%+({\hat T}_{T}^{(+)} + {\hat T}_{T}^{(-)})\,y^\pi_{2\Lambda}\big]
+\frac{\gA^2 }{8 \fpi^2} {\bar c}_{\Delta}
 \left[\frac13\TtimesS{\bar y}^\pi_{0\Lambda}
   -\frac{1}{2}\TtimesT y^\pi_{2\Lambda}\right]
  \hat j_1(qR)
\no &&-\frac{1}{16 f_\pi^2}{\bar N}_{WZ}\tau_1\cdot\tau_2
  \left[\vs_{+}{\bar y}^\pi_{0\Lambda}
+(3\hatr\hatr\cdot\vs_{+}-\vs_{+}) y^\pi_{2\Lambda}\right]
 \hat j_1(qR).
\label{vmu1piC}\ee
The values of the LECs,
$(\bar c_\omega,\ \bar c_\Delta,\ {\bar N}_{WZ})$,
should in principle be
fixed either by solving the underlying theory, QCD,
or by fitting to suitable experimental observables.
Since this has not yet been done,
we adopt here the estimates given in
Ref.~\cite{excV,isoS} based on the
resonance saturation assumption
and the Wess-Zumino action,
$(\bar c_\omega,\ \bar c_\Delta,\
{\bar N}_{WZ})\simeq (0.1021,\ 0.1667,\ 0.02395)$.
Further discussion on the use of
these estimates will be given later in the text.

The ``fixed-term" contributions,
$\vmu_{ij}^{1\pi:fixed}$,
represent vertex corrections to the {\em soft}-one-pion-exchange
and
fixed completely by Lorentz covariance. They might
be viewed as relativistic corrections to the $\pi NN$ and
$A^\mu_{\rm em} \pi NN$ vertices.
These corrections can be obtained conveniently
by performing the Foldy-Wouthuysen transformation of
the relativistic Lagrangian, and the resulting
vertex functions read

\noindent
$\bullet$ $\pi(k)+N(p)\rightarrow N(p')$ vertex : \bea &
&\frac{g_A}{2f_\pi}\tau^a \Big\{
     \vs\cdot\vk-\frac{k^0}{2m}\vs\cdot(2\vp+\vk)
\no & &  +\frac{1}{8m^2}[4\vp\cdot\vk\vs\cdot\vp-4\vp^2\vs\cdot\vk
     -2\vp\cdot\vk\vs\cdot\vk+2\vk^2\vs\cdot\vp-\vk^2\vs\cdot\vk]\Big\}.
\eea
\noindent
$\bullet$ $\pi^a(k)+N(p)\rightarrow N(p')+A_\mu^{\rm em}(q)$
vertex : \bea & &-i\frac{g_A}{2f_\pi}\epsilon^{3ag}\tau^g\Big\{
\vs\left(1-\frac{1}{8m^2}\left[
  4\vp^2-4\vp\cdot(\vq-\vk)+2\vk^2-2\vk\cdot\vq+\vq^2\right]\right)
\no
&&\qquad\quad+\frac{1}{4m^2}\left[ \vp \vs\cdot\vp' + \vp'
\vs\cdot \vp +\frac12 \vk\vs\cdot(2\vq-\vk) -\frac12
\vq\vs\cdot\vk + i\vk\times\vp\right]\Big\} \no &
&+\frac{g_A}{2f_\pi}(\tau^a+\delta^{3a})\Big\{\frac{k_0}{2m}\vs
+\frac{1}{4m^2}\left[
\vk\times(\vbp\times\vs)+\vs\times(\vbp\times\vk)
+i\vq\times\vk\right]\Big\}. \eea
The expressions of the M1 operator corresponding to the fixed terms
are quite lengthy and will be reported elsewhere~\cite{elsewhere}.
The fixed
terms containing the nucleon momentum operators make the calculation
highly involved.

The two-pion-exchange diagrams shown
in Fig.\ref{fig:N3LO-CT} give rise to
\be \vmu_{12}^{2\pi} &=& \frac{1}{128\pi^2
\fpi^4}\left[\left(\TplusS-\TminusS\right)L_S
              +\left(\TplusT-\TminusT\right)L_T\right]
   \hat j_1(qR)
\no &-& \frac{1}{256\pi^2 f_\pi^4}  \ttz \vR \times \hatr
\frac{d}{dr} L_0 \hat j_0(qR)
\,, \label{2pi}\ee
%&&- i \vq \times\left[({\hat T}_{S}^{(+)}-
%{\hat T}_{S}^{(-)} ) L_S + ({\hat T}_{T}^{(+)}- {\hat
%T}_{T}^{(-)}) L_T\right] \Big\} \label{2pi}\eea
where
\bea L_S &=& -\frac{g_A^2}{3} r \frac{d}{dr} K_0  +
\frac{g_A^4}{3} \big[ 4 K_1 -2 K_0+r\frac{d}{d r} (K_0 + 2 K_1)
\big], \no L_T&=&  \frac{g_A^2}{2}  r\frac{d}{d r} K_0 +
\frac{g_A^4}{2} \big[ 4 K_{\rm T} - r \frac{d}{d r} (K_0 + 2 K_1)
\big], \no L_0 &=&2 K_2 + g_A^2 (8K_2+2K_1+2K_0) \no && \ \ \
        -g_A^4 (16 K_2+5 K_1+5 K_0)+ g_A^4 \frac{d}{dr} (r K_1).
\eea
The loop functions $K's$ are defined in Ref. \cite{hep,excV}.

Finally there are contact-term contributions of the form \be
\vmu^{CT}_{12} &=& \frac{1}{2m_p}
         [g_{4S}(\vs_1+\vs_2)+g_{4V}T_S^{(\times)}]
         \delta^{(3)}_\Lambda(\vr) \hat j_0(qR),
\label{CT}\ee
where $g_{4S}=m_p g_4$ and $g_{4V}=- m_p
(G_A^R+\frac{1}{4}E_T^{V,R})$;
$g_4$, $G_A^R$ and $E_T^{V,R}$ are the coefficients of the
contact terms introduced in Refs.~\cite{excV,isoS}.
%A highly noteworthy
A noteworthy point is that,
after removing redundant terms,
there are only two independent contact-terms
relevant for the M1 operator up to \nlo3.
This reduction of the effective number of counter
terms is due to Fermi-Dirac statistics;
a similar reduction
has been noticed for the Gamow-Teller operator~\cite{hep},
where only one linear
combination of LECs needs to be retained.

%%%%%%%%%%%%%%%%%%%%%%%%%%%%%%%%%%%%%%%%%%%%%%%%%%%%%%%%%%%%%%%%%%%%%%%%%
\section{Results}
\indent

We now use the M1 operators described above to
calculate the magnetic moments of $\He3$, $\H3$ and $\H2$
as well as the rate of radiative capture
of a thermal neutron on a proton, $np\to d\gamma$.
The total $np\to d\gamma$ cross section at threshold reads
\be
\sigma_{np} =\frac{q^3}{v_n} \mu_N^2 \, M_{np}^2 \ee with \be
M_{np}= 2 m_p\, \int \! d^3 \vr \,
 \psi_{d,0}^\dagger(r) \vmu^z(\vq) \psi_{np}(r),
\ee
where $v_n\!=\!2,200\ \mbox{m/s}$
denotes the neutron velocity in
the lab frame,
$\psi_{d,0}(r)$ the deuteron wave function with the  spin
component $J_z=0$. The $\psi_{np}(r)$ denotes the
spin-singlet $np$ scattering wave function obtained by applying
the normalization condition $\lim_{r\to \infty} u_{np}(r) = r -
a_s$, where $a_s$ is the spin-singlet $np$ scattering length.
The experimental value of the
radiative $np$ capture cross section is
$\sigma_{np}^{exp}=332.6\pm 0.7$ mb~\cite{npexp},
which corresponds to $M_{np}^{exp}= 410.2\pm
0.4\ \mbox{fm}^{3/2}$.

A realistic nuclear Hamiltonian is constructed using
a high-precision phenomenological potential.
We consider here the Av18 bi-nucleon potential~\cite{Av18}
with/without the Urbana IX (U9) tri-nucleon force
(3NF)~\cite{UIX}, and we also consider
the Argonne Av14 bi-nucleon
potential with/without the Urbana VIII tri-nucleon force (U8).
We remark that, from the HBChPT point of view,
3NF is \nlo3 compared to the leading
nucleon-nucleon interactions, and hence
should be included in our \nlo3 calculation.

The variational Monte Carlo (VMC)
technique~\cite{UIX,Wiringa91} is used to solve
the 3-body Schroedinger equation
in order to obtain wave functions of $\H3$ and $\He3$.
MEEFT is based on the assumption that the adopted
nuclear wave functions are exact
to the order under consideration.
This presupposes (i) the validity of
using the high-precision phenomenological potential
instead of an EFT-based potential,
and (ii) the sufficient accuracy of the method used
for solving the A-body Schroedinger equation
for a given phenomenological A-body Hamiltonian.
The first point is a fundamental issue
that awaits further detailed studies.
As for the second point, we remark that
the long-range behavior of the wave functions
can significantly affect the calculated
nuclear transition matrix elements,
and hence one should avoid using schematic
wave functions.
Unfortunately, it is difficult to fully quantify the uncertainty
related to the wave functions used in this work.
One of indirect measures for the accuracy
of the adopted potentials
and the VMC technique
is the binding energies
%-- which may be viewed as the most relevant quantity
%   of low-energy nuclear systems --
of $\H3$ and $\He3$.
We give in Table~\ref{VMCtable} the results
obtained in VMC for the Av18 and Av18+U9 potentials;
also shown in the table are the results obtained
by solving the Faddeev
equations~\cite{Faddeev,Thesis,CarbLaz_PRC04}.
Note that the use of 3NF is imperative in
order to reproduce the binding energies of $\H3$ and $\He3$.
Later in the text we shall give comparison of the results
for the M1 observables calculated with the Av18(+U9) potential
and the Av14(+U8) potential, and this comparison
provides another indirect measure for the stability of
our results against changes in the input potential.

\begin{table}
\caption{\label{VMCtable} The binding energies (in MeV)
of $\H3$ and $\He3$
calculated in the VMC method for the Av18 and Av18+U9 potentials.
For comparison, we also give in the square brackets
the results obtained by solving the Faddeev equations. }
\begin{center}
\begin{tabular}{c|c|c|c}
\hline
          & Av18 &  Av18+U9 & Exp \\
\hline
BE($\H3$) &7.35(1) [7.61]  & 8.24(1) [8.47] & 8.48  \\
BE($\He3$)&6.59(1) [6.91]  & 7.48(1) [7.74] & 7.72  \\
\hline
\end{tabular}
\end{center}
\end{table}

The cutoff $\Lambda$ has a physical meaning,
and its choice is not arbitrary~\cite{cutoff}.
%\syh{The validity of the
%cutoff larger than the $\Lambda_\chi$ is a debating
%subject\cite{cutoff}.
%}
Thus $\Lambda$ should be
smaller than the masses of the vector mesons that have
been integrated out.
Meanwhile, since the pion is an explicit degree of freedom in our
scheme, $\Lambda$ should be much larger than the pion mass in
order to ensure that all pertinent low-energy contributions are
properly included. In the present work, we consider
$\Lambda=$ 500, 600, 700 and 800 MeV as representative values.

\begin{table}
\caption{\label{VMCtable:wrt-cutoff}
 Magnetic moments of
$\H3$ and $\He3$ in units of the nuclear magneton calculated
for the Av18+U9 interactions.
Also listed are the values of
$g_{4s}$ and $g_{4v}$ which, for
a given value of $\Lambda$, reproduce the experimental values of
$\mu(\H2)$ and $M_{np}$.
The parenthesized numbers represent the Monte Carlo statistical errors,
while the entries in the square
brackets correspond to the calculations in which the contact term
contributions are ignored.}
\begin{center}
\begin{tabular}{r|r|r|rr}
\hline
$\Lambda$ [MeV]  & $\mu(\H3)$ & $\mu(\He3)$ & $g_{4s}$ & $g_{4v}$  \\
\hline
500& $3.034(13) \ [2.883]$&  $-2.196(13) [-2.074]$  & 0.5786  & 2.8790 \\
600& $3.035(13) \ [2.944]$&  $-2.198(13) [-2.120]$  & 0.2995  & 1.9567\\
700& $3.036(13) \ [2.988]$&  $-2.199(13) [-2.150]$  & -0.0202 & 1.2954\\
800& $3.037(13) \ [3.019]$&  $-2.200(13) [-2.168]$  & -0.3965 & 0.7882\\
\hline
\end{tabular}
\end{center}
\end{table}
For a given value of $\Lambda$, we adjust
$g_{4s}$ and $g_{4v}$ to reproduce
the experimental values of
$\mu(\H2)=0.8574\mu_N$ and $\sigma_{np}^{exp}$ (or
equivalently $M_{np}^{exp}$).
We remark that the fitted values of $g_{4s}$ and
$g_{4v}$ depend on a particular choice of the potential model
used to calculate the wave functions as well as
the cutoff parameter.
In Table~\ref{VMCtable:wrt-cutoff}, we show
our predictions for the $\H3$ and $\He3$ magnetic moments
obtained with the Av18+U9 potential for
the four representative values of $\Lambda$;
also shown are the values of $g_{4s}$ and $g_{4v}$
optimized for each value of $\Lambda$.
To highlight the roles of the contact terms,
we include in the table the results corresponding
to cases for which the contact terms are artificially dropped;
see the entries in the square brackets.
Table~\ref{VMCtable:wrt-cutoff} indicates
that the results of the full calculation are
almost independent of $\Lambda$,
whereas those obtained without the contact-term
contributions show pronounced cutoff dependence.

\begin{table}
\caption{\label{VMCAv18U9} The magnetic moments
of $\He3$, $\H3$, $\H2$ (in units of $\mu_N$) and the
$np\rightarrow d\gamma$
matrix elements (in units of $\mbox{fm}^{3/2}$)
calculated with the Av18+U9 potential
for $\Lambda=600$ MeV.}
\begin{center}
\begin{tabular}{l|r|r|r|r}
\hline
                   &$\mu(\H2)$&$M_{np}$&$\mu(\H3)$&$\mu(\He3)$ \\
\hline
LO: 1B                & 0.8469& 393.1& 2.585&-1.774\\
\hline
NLO: $1\pi$         & 0.0000&  8.7&  0.205& -0.205\\
\hline
\nlo2: 1B $1/m^2$   &-0.0069&  -0.1&-0.018&-0.007\\
\hline
\nlo3: $1\pi C$     & 0.0077&  4.4& 0.133&-0.116\\
\nlo3: fixed term   & 0.0044& -0.2&-0.003& 0.014\\
\nlo3: $2\pi$       & 0.0000&  1.5& 0.043&-0.043\\
\nlo3: $g_{4s}$ term& 0.0053&  0.0& 0.007& 0.007\\
\nlo3: $g_{4v}$ term& 0.0000&  2.9& 0.085&-0.085\\
\nlo3 total         & 0.0174&  8.6& 0.265&-0.223\\
\hline
total               & 0.8574&410.2& 3.035&-2.198\\
\hline
experiment          &0.8574&410.2(4)&2.979&-2.128\\
\hline
\end{tabular}
\end{center}
\end{table}

Table~\ref{VMCAv18U9} shows the contributions of
the individual M1 operators to the magnetic moments of $\He3$,
$\H3$ and $\H2$ as well as the matrix element
for $np\rightarrow d\gamma$,
calculated with the Av18+U9 potential
for $\Lambda=600\ \MeV$.
{}The table indicates that the NLO $1\pi$ contribution
is rather small, comparable to the total \nlo3 contribution.
Substantial cancellation
among the various terms contributing to $\vmu^{1\pi}$
is responsible for this feature.
Thus
the $\vmu^{1\pi}$ contribution to $M_{np}$,
$8.7$ in units of ${\mbox{fm}}^{3/2}$,
can be decomposed as
$8.7=(19.4-14.4+4.0)-0.3$,
where the first three terms come from the first line
of eq.(\ref{1pi}), and the fourth term
from the second line.

%\begin{table}
%\caption{\label{VMCtable:wrt-pot:old} {\bf To be removed.} Magnetic moments of
%$\H3$ and $\He3$ calculated for various potentials but with the fixed
%cutoff value $\Lambda=600$ MeV. The values of $g_{4s}$ and
%$g_{4v}$ that reproduce the experimental values of $\mu(\H2)$ and
%$M_{np}$ are also listed. }
%\begin{center}
%\begin{tabular}{r|r|r|rr}
%\hline
%potential  & $\mu(\H3)$ & $\mu(\He3)$ & $g_{4s}$ & $g_{4v}$  \\
%\hline
%Av18+U9  & $3.035(13) $ & $-2.198(13)$ & $ 0.300(6)$ & $1.957(289)$ \\
%Av18     & $3.040(13) $ & $-2.202(13)$ & $ 0.300(6)$ & $1.957(289)$ \\
%Av14+U8  & $3.052(13) $ & $-2.214(13)$ & $ 0.391(7)$ & $2.247(310)$\\
%Av14     & $3.024(12) $ & $-2.180(12)$ & $ 0.391(7)$ & $2.247(310)$\\
%\hline
%\end{tabular}
%\end{center}
%\end{table}
%
\def\psp{\phantom{$(12)$}}
\def\pspI{\phantom{$(0)$}}
\def\pspts{\phantom{This work }}
\begin{table}
\caption{\label{VMCtable:wrt-pot}
The $A=3$ magnetic moments
(in units of $\mu_N$) calculated for $\Lambda=600$ MeV.
The last column gives the
values of $g_{4s}$ and
$g_{4v}$ fitted to reproduce the experimental values of
$\mu(\H2)$ and $M_{np}$.
The bottom three rows show the SNPA results and
the experimental data. }
\begin{center}
\begin{tabular}{l|c|c|cc} \hline
   & $\mu(\H3)+\mu(\He3)$ & $\mu(\H3)-\mu(\He3)$ & $g_{4s}$ & $g_{4v}$  \\
\hline
This work (Av18+U9)  & $0.838(0) $ & $5.233(25)$ & $ 0.300(6)$ & $1.96(29)$ \\
\pspts(Av18)     & $0.838(0) $ & $5.242(26)$ & $ 0.300(6)$ & $1.96(29)$ \\
\pspts(Av14+U8)  & $0.838(0) $ & $5.266(33)$ & $ 0.391(7)$ & $2.25(31)$\\
\pspts(Av14)     & $0.844(0) $ & $5.204(30)$ & $ 0.391(7)$ & $2.25(31)$\\
\hline
SNPA I \cite{full-old}        & 0.828\pspI & 5.078\psp & & \\
SNPA II\cite{Marcucci:2005zc} & 0.884\pspI & 5.114\psp & &\\
\hline
Experiment\cite{mmexp}        & 0.851\pspI    & 5.107\psp  & &\\
\hline
\end{tabular}
\end{center}
\end{table}
Finally, in Table~\ref{VMCtable:wrt-pot}
we compare the results for the different
choices of the nuclear potentials;
the table also gives the results of the
SNPA calculations taken from
Refs.~\cite{full-old,Marcucci:2005zc},
as well as the experimental data~\cite{mmexp}.
The numbers in the parentheses represent statistical errors
of our VMC calculation.
We remark that the uncertainty in the experimental value
of $\sigma_{np}$, which affects the determination of
$g_{4s}$ and$g_{4v}$, causes $\sim 0.5\ \%$
uncertainty in $\mu(\H3)$-$\mu(\He3)$.
Our results are found to be almost independent of the interaction
model and the cutoff values used, satisfying the general
tenet of EFTs that the model- and cutoff-dependence should be
of higher order than the order to which calculation is performed,
i.e., \nlo3 for the present case. This
potential-independence is also consistent with the notion of the
$V_{low-k}$ \cite{Vlowk}, which dictates that all the
high-precision phenomenological potentials become universal if the
cutoff is lowered to $~2\ {\mbox{fm}}^{-1}$. Physically,
this reflects the fact that
the differences among various high-precision potentials
lie only in the high-energy region that can be renormalized away
leaving little effect for low-energy dynamics, provided that the
renormalization procedure is correctly done.

Comparison of the results with and without
the tri-nucleon forces (3NF) in Table~\ref{VMCtable:wrt-pot}
indicates that the role of 3NF in the magnetic moments
is small; the largest effect seen for the Av14+U8 case
is of the level of 1\%.

Our no-parameter MEEFT calculation gives the values
of $\mu(\H3)+\mu(\He3)$ and $\mu(\H3)-\mu(\He3)$
which agree with the experimental values
at the 2-3 \% level.
This is slightly worse than the agreement reported
in the SNPA calculations.  We will come back to
this point in the next section.
We emphasize, however, that as far as the structure
of the M1 operators is concerned,
the present work gives a complete expression,
a feature that distinguishes MEEFT from SNPA.

\section{Discussion}
\indent

We have reported here the calculation of the
magnetic moments of $\H3$ and $\He3$
based on HBChPT.
All the M1 operators up to \nlo3
have been explicitly derived.
At \nlo3, two unknown parameters,
$g_{4s}$ and $g_{4v}$, enter
as the coefficients of contact terms.
Following the MEEFT strategy, we have fixed them
by imposing the renormalization condition
that the experimental values of the deuteron magnetic moment
and thermal neutron capture rate on proton
be reproduced at \nlo3.

Concerning the  $2\sim3\ \%$ discrepancy between
the values of
$\mu(\H3)+\mu(\He3)$ and $\mu(\H3)-\mu(\He3)$
obtained in our MEEFT calculation
and the corresponding experimental values,
there are a few points to be discussed.
The first is the level of accuracy
of the $A=3$ wave functions obtained in the VMC method.
The indication that the VMC wave functions deviate slightly
from the accurate wave functions
is already visible in Table~\ref{VMCtable},
which shows that the binding energies of the $A=3$ systems
calculated in VMC do not quite agree with those
obtained in the Faddeev calculation~\cite{Faddeev,Thesis,CarbLaz_PRC04}.
An attempt is in progress~\cite{elsewhere}
to improve the present work
with the use of {\it ab initio} wave functions obtainable
in the Faddeev method
and to extend our formalism to the
$\He3+n\rightarrow \He4+\gamma$ reaction.

The second point is higher order contributions.
%---TSP
%We note that,
As can be seen in Table~3,
the difference between theory and experiment
is only about 30~\% of \nlo3 contributions,
for both isovector and isoscalar channel.
Given the HB$\chi$PT expansion parameter,
this suggests that \nlo4 effects of natural size can remedy
the discrepancy.
%---
%For example, the discrepancy can be easily resolved
%by assuming the \nlo4 contributions
%to be as large as one-third
%of the \nlo3 contributions.
%
Among the \nlo4 contributions,
we expect the three-body (3B) currents
-- which appear first at \nlo4
and hence have not been studied here--
to play the most important role
for the following reasons.
%Due to the renormalization procedure for the coefficients
%of the contact terms (CT),
We note that a substantial portion of the higher-order two-body contributions
%(other than CTs)
can be effectively absorbed into the
renormalization of the coefficients of the contact terms;
this is particularly true for short-ranged contributions.
To illustrate this aspect,
we consider the role of the $2\pi$ contribution
in our \nlo3 calculation
and compare the results of two
``$2\pi$-contribution-less" calculations.
In the first case, the $2\pi$ terms are simply dropped
without any other accompanying changes,
whereas in the second case
the values of  $g_{4s}$ and $g_{4v}$ are readjusted
to reproduce the experimental values
of $\mu(\H2)$ and $\sigma_{np}$
({\em without} the $2\pi$ contributions).
We expect that the change in the net results
for the second case,
which may be viewed as
an {\em effective} $2\pi$ contribution,
is much smaller
than the $2\pi$ contribution estimated from the first case.
We have verified this feature numerically.
From this example,  we may expect that
the {\em effective} \nlo4 2B contributions are small.
The situation is however completely different for
the 3B contribution
in the M1 operator,
($\vmu_{\rm 3B}$).
Since $\vmu_{\rm 3B}$ appears
in the $A=3$ systems but not in the $A=2$ systems,
its contribution cannot be absorbed into the two-nucleon contact terms.
The higher order calculation up to \nlo4 is
relegated to future work.
%\syh{And the fully consistent
%treatment in the EFT formalism such as \cite{Krebs:2004st}
%will be valuable.}

Finally, we discuss the uncertainty
related to the values of the LECs,
$\bar c_\omega$, $\bar c_\Delta$ and ${\bar N}_{WZ}$,
which appear in $\vmu_{12}^{1\pi C}$, eq. (\ref{vmu1piC}).
The $\vmu_{12}^{1\pi C}$ contributions themselves
are comparable in size to the difference
between the present calculation and the experimental data,
as can be seen in Table~\ref{VMCAv18U9}.
However, due to the above-mentioned renormalization procedure,
the {\em effective}
$\vmu_{12}^{1\pi C}$ contributions
are expected to be small.
We have checked this by turning off $\vmu_{12}^{1\pi C}$
and re-performing the renormalization procedure,
and found that the {\em effective}
$\vmu_{12}^{1\pi C}$ contributions are negligible.
If on the other hand we naively treat these LECs as free parameters
and adjust them to reproduce the experimental values,
we are led to very large values of the LECs,
which strongly violates the naturalness condition.

%%%%%%%%%%%%%%%%%%%%%%%%%%%%%%%%%%%%%%%%%%%%%%%%%%%
\section*{Acknowledgements}
We would like to acknowledge invaluable discussions with, and
support from, Professors Kuniharu Kubodera and Mannque Rho with
whom this work was initiated.
YH and TSP are grateful for helpful discussions with Daniel Phillips,
Young-Man Kim.
The work of TSP was supported
in part by grant No. R01-2006-10912-0 from the Basic
Research Program of the Korea Science \&
Engineering Foundation and in part by
Asia Pacific Center for Theoretical Physics (APCTP).

%==============================================================

%----------------------------------------------------

\end{document}